\documentstyle[12pt,epsfig]{article}

\advance\textheight by \topskip
\newcommand{\be}{\begin{equation}}
\newcommand{\ee}{\end{equation}}
\newcommand{\br}{\begin{eqnarray}}
\newcommand{\bea}{\begin{eqnarray}}
\newcommand{\beanon}{\begin{eqnarray*}}
\newcommand{\er}{\end{eqnarray}}
\newcommand{\eea}{\end{eqnarray}}
\newcommand{\eeanon}{\end{eqnarray*}}
\newcommand{\ba}{\begin{array}}
\newcommand{\ea}{\end{array}}
\newcommand{\bi}{\begin{itemize}}
\newcommand{\ei}{\end{itemize}}
\newcommand{\bn}{\begin{enumerate}}
\newcommand{\en}{\end{enumerate}}
\newcommand{\bc}{\begin{center}}
\newcommand{\ec}{\end{center}}

\newcommand{\ar}{\rightarrow}

\newcommand{\Dir}{\kern -6.4pt\Big{/}}
\newcommand{\Dirin}{\kern -10.4pt\Big{/}\kern 4.4pt}
\newcommand{\DDir}{\kern -7.6pt\Big{/}}
\newcommand{\DGir}{\kern -6.0pt\Big{/}}
\def\sm{\ifmmode{{\cal {SM}}}\else{${\cal {SM}}$}\fi}
\def\mssm{\ifmmode{{MSSM}}\else{MSSM}\fi}
\def\MH{\ifmmode{{M_{H}}}\else{${M_{H}}$}\fi}
\def\Mh{\ifmmode{{M_{h}}}\else{${M_{h}}$}\fi}
\def\MA{\ifmmode{{M_{A}}}\else{${M_{A}}$}\fi}
\def\MWpm{\ifmmode{{M_{W^\pm}}}\else{${M_{W^\pm}}$}\fi}
\def\MHpm{\ifmmode{{M_{H^\pm}}}\else{${M_{H^\pm}}$}\fi}
\def\Wpm{\ifmmode{{{W^\pm}}}\else{${{W^\pm}}$}\fi}
\def\Hpm{\ifmmode{{{H^\pm}}}\else{${{H^\pm}}$}\fi}
\def\tb{\ifmmode{\tan\beta}\else{$\tan\beta$}\fi}
\def\ctb{\ifmmode{\cot\beta}\else{$\cot\beta$}\fi}
\def\ta{\ifmmode{\tan\alpha}\else{$\tan\alpha$}\fi}
\def\cta{\ifmmode{\cot\alpha}\else{$\cot\alpha$}\fi}
\def\tba{\ifmmode{\tan\beta=1.5}\else{$\tan\beta=1.5$}\fi}
\def\tbb{\ifmmode{\tan\beta=30.}\else{$\tan\beta=30.$}\fi}
\def\cab{\ifmmode{c_{\alpha\beta}}\else{$c_{\alpha\beta}$}\fi}
\def\sab{\ifmmode{s_{\alpha\beta}}\else{$s_{\alpha\beta}$}\fi}
\def\cba{\ifmmode{c_{\beta\alpha}}\else{$c_{\beta\alpha}$}\fi}
\def\sba{\ifmmode{s_{\beta\alpha}}\else{$s_{\beta\alpha}$}\fi}
\def\ca{\ifmmode{c_{\alpha}}\else{$c_{\alpha}$}\fi}
\def\sa{\ifmmode{s_{\alpha}}\else{$s_{\alpha}$}\fi}
\def\cb{\ifmmode{c_{\beta}}\else{$c_{\beta}$}\fi}
\def\sb{\ifmmode{s_{\beta}}\else{$s_{\beta}$}\fi}
\def\Ord{\buildrel{\scriptscriptstyle <}\over{\scriptscriptstyle\sim}}
\def\OOrd{\buildrel{\scriptscriptstyle >}\over{\scriptscriptstyle\sim}}
\def\sm{\ifmmode{{\cal {SM}}}\else{${\cal {SM}}$}\fi}
\def\mt{\ifmmode{{m_{t}}}\else{${m_{t}}$}\fi}
\def\MH{\ifmmode{{M_{H}}}\else{${M_{H}}$}\fi}
\def\MWpm{\ifmmode{{M_{W^\pm}}}\else{${M_{W^\pm}}$}\fi}
\def\Wpm{\ifmmode{{{W^\pm}}}\else{${{W^\pm}}$}\fi}
\def\pl #1 #2 #3 {{\it Phys.~Lett.} {\bf#1} (#2) #3}
\def\np #1 #2 #3 {{\it Nucl.~Phys.} {\bf#1} (#2) #3}
\def\zp #1 #2 #3 {{\it Z.~Phys.} {\bf#1} (#2) #3}
\def\pr #1 #2 #3 {{\it Phys.~Rev.} {\bf#1} (#2) #3}
\def\prep #1 #2 #3 {{\it Phys.~Rep.} {\bf#1} (#2) #3}
\def\prl #1 #2 #3 {{\it Phys.~Rev.~Lett.} {\bf#1} (#2) #3}
\def\mpl #1 #2 #3 {{\it Mod.~Phys.~Lett.} {\bf#1} (#2) #3}
\def\rmp #1 #2 #3 {{\it Rev. Mod. Phys.} {\bf#1} (#2) #3}
\def\xx #1 #2 #3 {{\bf#1}, (#2) #3}

\begin{document}
\tolerance=100000
\thispagestyle{empty}
\setcounter{page}{0}

\begin{flushright}
{\large Cavendish--HEP--97/13}\\
{\rm August 1997\hspace*{.5 truecm}}\\ 
\end{flushright}

\vspace*{\fill}

\begin{center}
{\large \bf Charged Higgs scalar production\\[0.25cm]
in single-top mode (and other)\\[0.25cm] 
at future $ep$ colliders\\[0.30cm]
in the Minimal Supersymmetric Standard Model}\\[2.cm]
{\large Stefano Moretti$^1$ and Kosuke Odagiri\footnote{E-mail: 
moretti, odagiri@hep.phy.cam.ac.uk}}\\[.3 cm]
{\it Cavendish Laboratory, University of Cambridge,}\\
{\it Madingley Road, Cambridge, CB3 0HE, United Kingdom.}\\[1cm]
\end{center}

\vspace*{\fill}

\begin{abstract}{\small\noindent 
We study charged Higgs boson production at future electron-proton colliders 
in the framework of the Minimal Supersymmetric Standard Model.
We focus our attention to the case of single-top production and decay
through the channel $t\rightarrow bH^\pm$ and of vector-scalar fusion 
via $W^{\pm*} \Phi^*\rightarrow H^\pm$ (where $\Phi=H,h$ and
$A$). We consider the signature
$H^\pm\rightarrow \tau\nu_\tau$ and compare it to the irreducible 
background from Standard Model interactions. For $M_{H^\pm}\Ord  m_t$,
the $H^\pm$ signal is accessible through lepton universality breaking 
if $M_A\Ord 100-120$ GeV at both low and large values of $\tb$. Furthermore,
although the bulk of the production cross section comes from single-top
events, a sizable contribution due to vector-scalar-scalar interactions
should be observable at large $\tb$, this possibly 
offering some insights into the 
structure of the scalar sector of the theory.
The possibility
of the CERN collider running in the LEP$\oplus$LHC mode is considered in 
detail.}\end{abstract}

\vspace*{\fill}
\newpage

\section*{1. Introduction and motivation} 

A charged Higgs boson is a building block of two Higgs Doublet Models
(2HDMs), including the Supersymmetry (SUSY) version, as well as of 
Technicolour (TC) theories \cite{guide}. Conversely, such a particle does not 
belong to the spectrum of the Standard Model (SM). Therefore, its detection 
would be an unequivocal signal of New Physics. 

To date, a lower limit on the value of its mass has been set by the LEP2 data,
yielding $\MA\OOrd60$ GeV (for $\tb\OOrd1$) \cite{Alimit}, through the 
(tree-level) relation $\MHpm^2=\MWpm^2+\MA^2$, so that $\MHpm\OOrd100$ GeV. 
From arguments related to the request of unitarity
of the underlying theory one should expect the upper limit being in the 
TeV region \cite{unitarity}. Therefore, the mass range allowed for the existence
of charged Higgs bosons is  indeed vast. Though, if one confines oneself to
the case of the Minimal Supersymmetric Standard Model (MSSM) --as we do in the 
present paper-- the decay spectrum of such scalars is rather simple. 
If one further assumes that the mass scale of the Supersymmetric partners
of ordinary matter is above the $H^\pm$ mass, then only two modes 
dominate the decay phenomenology of the charged Higgs boson.
Their reciprocal relevance is dictated by the interplay between the 
Higgs and top masses. If
$\MHpm\Ord m_t$, then the branching ratio (BR) $\mbox{BR}(H^\pm\ar
\tau\nu_\tau)$ is the largest (around
$98\%$, for $\tan\beta>2$) and depends only slightly on the $\beta$ angle.
When $M_{H^\pm}\OOrd m_t$,
the $H^\pm\ar bt$ decay mode is the only accessible channel 
(with a BR of  practically $100\%$ at all $\tb$'s).
The mode $H^\pm\ar hW^\pm $ 
can be relevant only at small values of $\tb$ and in a very narrow
mass region right below the $bt$ decay threshold \cite{ioejames}. 

As for the production mechanisms of charged Higgs bosons at 
colliders, it is likely that one will have to wait for the advent of the
future generation of high energy accelerators,
in order to detect such particles
(see Ref.~\cite{guide} for a review). In fact, 
at LEP2, the huge irreducible background
in $e^+e^-\ar W^+W^-$ events renders the signal
$e^+e^-\ar H^+H^-$ \cite{wrk41,wrk112} very hard to extract.
In addition, after the recent limit on $\MA$, the discovery potential of such 
a machine is rather poor, being confined to a tiny window of a few GeV and
only if the CERN $e^+e^-$ machine will reach the energy 
$\sqrt s_{ee}=205$ GeV \cite{eeHH}, which was considered in the context of 
the 1995 LEP2 Workshop \cite{lep2w}.

Since the fact that the charged Higgs mass falls right within the discovery
potential of LEP2 is clearly matter of luck,
it is far-seeing to look at the case of future machines. At the CERN
Large Hadron Collider (LHC), the 
$H^\pm$ scalar of the MSSM is expected 
to be copiously produced in top quark decays
$t\ar bH^\pm$, provided that
$m_t\OOrd M_{H^\pm}$ and the value of $\tan\beta$
is low or high enough\footnote{The minimum of the $t\ar bH^\pm$
decay rate is at about $\tan\beta=6$.}. Top quarks
are produced in $t\bar t$ pairs, with a large cross section (we assume
$m_t=175$ GeV) \cite{ATLAS,CMS} and the
charged Higgs boson is searched for
by means of the leptonic signature $H^\pm\ar\tau\nu_\tau$.
Since neutrinos prevent one from reconstructing the
Higgs mass from the momenta of its decay products, the existence
of $H^\pm$ signals in the data can be inferred only from an excess
of $\tau$ production with respect to what is predicted in the SM
(the {\sl lepton universality breaking} signal).

In contrast, if $M_{H^\pm}\OOrd m_t$  the 
chances of detection at the LHC are very much reduced. In fact, not only 
the known production mechanisms of MSSM charged Higgses yield small
cross sections, but also the lack of a clean signature 
contributes to make the signal very poor and 
overwhelmed by the ordinary 
QCD background. On the one hand, 
only the subprocesses $bg\ar tH^\pm$ \cite{guide81}
and $bq\ar bq'H^\pm$  \cite{ioeKosuke} can be of 
some help if $\MHpm\Ord300-400$ GeV \cite{guide} (and
large values of \tb\ in the second case, too). On the other hand, in order to 
extract the signal from the huge $W^\pm$~+~jets noise typical of hadron-hadron 
colliders \cite{guide84}, one would really need to reconstruct the Higgs mass 
resonance through the decay chain $H^\pm\ar bt\ar b\bar b W^\pm\ar b\bar b jj$ 
(where $j$ represents a jet from the $W^\pm$ decay), procedure
which relies on very high $b$-tagging performances and jet resolution
(and dedicated `tricks'
to trigger the `stiff lepton' from the primary top in $tH^\pm$ production 
\cite{guide81})\footnote{Note that if SUSY decays of charged Higgs 
bosons are allowed, several novel signatures (mainly 
involving charginos and neutralinos \cite{guide23,guide24,guide132})
could be exploited \cite{guide}, though their phenomenology is at 
present very much SUSY-parameter dependent for being of experimental concern. 
This issue is however beyond the scope of this paper.}.     

At the Next Linear Collider (NLC) \cite{ee500}--\cite{JLC},
$H^\pm$ detection looks easy,
though the discovery potential of such a
machine is far for being decisive \cite{sopczak}. Once again the crucial 
point
is the heavy mass range. Being an $e^+e^-$ collider, 
it can boast the advantage
of a much smaller QCD noise (as compared to hadron-hadron machines), however,
in this case is the maximum centre-of-mass
(CM) energy which sets the upper limit on the detectable 
$H^\pm$ mass. For a $\sqrt s_{ee}=500$ GeV NLC
\cite{ee500,ee500new}, one clearly cannot go beyond the value 
$\MHpm\approx220$ GeV, as the main production channel is 
$e^+e^-\ar H^+H^-$ \cite{wrk112}. Furthermore, the
$e^\pm\gamma$ and $\gamma\gamma$ running modes (using Compton back-scattered
laser photons \cite{Kozanecki,djm}) at the NLC do not improve the
prospects of MSSM charged Higgs detection \cite{eph29}, as the three viable
channels $e^\pm\gamma\ar e^\pm H^+H^-$, $e^\pm\gamma\ar\nu_e H^\pm A$ 
\cite{ioeph}
and $\gamma\gamma\ar H^+H^-$ \cite{eph33} only allow one to cover adequately
the intermediate Higgs mass range.
In this context, much higher CM energies \cite{JLC} would be more 
helpful, though the realisation of such designs is 
well into the next millenium.

Since the detection of heavy charged Higgs scalars of the MSSM is far from 
certain even after the end of the LHC and NLC era, it is particularly worthwhile
to assess already at present the discovery potential of these particles of 
other planned machines too. This will anyway be beneficial, 
whichever the outcome of the actual analyses is. In fact, these will either 
establish the impracticability of charged Higgs boson searches or, more
interestingly, provide a new experimental ground where to test the MSSM theory.
In particular, they could well extend the present coverage in mass and/or 
offer alternative production mechanisms of charged Higgs scalars, the latter 
involving new interactions other than the $t\ar bH^\pm$ decay (LHC) and the
QED-like vertex $\gamma\ar H^+H^-$ (NLC).

We turn our attention to the case of future electron(positron)-proton 
colliders, running with a CM energy in the TeV 
range\footnote{The Higgs discovery potential of the only $ep$ collider
operative at present, i.e., HERA at DESY, has been shown to be very 
poor \cite{HERA}. As for charged Higgses in the MSSM, the 
available production channels are via $\gamma\gamma$ \cite{abu32}, 
$\gamma g$ \cite{abu33} and $\gamma q$ fusion \cite{abu34}, all being 
significant only for very light scalar masses (strongly
disfavoured by the experiment).}. The physics of $ep$ colliders, in 
conjunction with the discussed possibility of their running in the $\gamma p$ 
mode \cite{gamma_p}, has been recently under renewed and active discussion
\cite{ankara}. A possible design was proposed and several
experimental simulations performed already in 1990
(during the Aachen LHC Workshop \cite{LHC}), 
for a LEP$\oplus$LHC \cite{LHC_phys} machine 
obtainable crossing one electron(positron) beam from LEP and a proton
one from the LHC \cite{verdier}. 
For a 100 GeV electron(positron) and a 7 TeV proton, 
the total energy in the frame of the colliding particles would be
$\sqrt s_{ep}\approx1.7$ TeV. Depending on the relative values of the 
electron(positron) and proton energy, the instantaneous luminosity should vary
in the range $(5\times10^{31}-4\times10^{32})$ cm$^{-2}$ s$^{-1}$
\cite{LHC_exp}. We convert
these values into 1 fb$^{-1}$ of integrated luminosity per annum, number that
we will adopt as a default in the forthcoming discussions.
The attractiveness of such a machine is
that it allows for a cleaner environment thanks to
the suppression of the
initial state QCD noise while
maintaining the collision energy at the TeV scale.

To our knowledge, no detailed study of MSSM charged Higgs boson production
at future $ep$ colliders exists in the literature, apart from a preliminary 
analysis carried out in Ref.~\cite{Cruz-Sampayo}. However, we do expect that
charged Higgs bosons of the MSSM can be abundantly produced in 
electron(positron)-proton collisions at the TeV scale. In particular,
it is the purpose of this paper to study the reaction (e.g., for the
case of a positron beam)
\be\label{bH}
e^+ b\ar \bar\nu_e bH^{+},
\ee
proceeding through the
two subprocesses 
\be\label{tbH}
e^+ b\ar \bar\nu_e t\ar \bar\nu_e bH^{+}
\ee
(i.e., single top production and decay)
and\footnote{Here
and in the following $\Phi$ refers collectively to the three
neutral Higgs scalars of the MSSM: $H,h$ and $A$.} 
\be\label{VSfusion}
e^+ b\ar \bar\nu_e b W^{\pm *}\Phi^{*}\ar \bar\nu_e bH^{+}
\ee
(i.e., vector-scalar fusion).
If one considers two-body fermion decays of the charged Higgs, then the graphs
contributing to
\be\label{proc}
e^+ b\ar \bar\nu_e bH^{+}\ar \bar\nu_e b f\bar f',
\ee
(where $ff'$ represents, e.g., $\tau^+\nu_\tau$ or $bt$) 
are those depicted in Fig.~1.

We are motivated to study this process following the results presented
in Ref.~\cite{ioeKosuke}, where the hadronic counterpart of process (\ref{proc})
was considered (i.e., $e^+\ar q$ and $\bar\nu_e\ar q'$, with $q^{(')}$
light quark). There, it was shown that $bq$ fusion could effectively help 
in increasing the chances of $H^\pm$ detection at the LHC, also above
the $H^\pm\ar bt$ decay threshold. This is due to three main reasons:
(i) a large content of $b$-quarks inside the proton at the LHC;
(ii) the strength of the Yukawa couplings of the neutral Higgs bosons
of the MSSM to the $b$-quarks increasing with the value of $\tan\beta$
(graph 2); (iii)
vertex tagging performances of the LHC detectors which are expected to
be ideal.

One should expect this channel to be similarly
effective also at a future $ep$ collider, as the large content of $b$ quark 
inside the scattered hadron is guaranteed by the TeV energy of the LHC
beam and the capabilities of the LHC vertex detector should be maintained
while running the CERN machine in the $ep$ mode. Along with the signal
(\ref{bH}) we will also study several SM-like `irreducible' backgrounds, on the
same footing as in Ref.~\cite{ioeKosuke}\footnote{For the time being,
we neglect considering $H^\pm$ signals via top production and decay 
in double mode, through $\gamma^* g\ar t\bar t$, as the corresponding cross
section is, at the TeV scale and for $m_t=175$ GeV, a factor of 100 smaller 
than that of single-top \cite{LHC_top,epSM}.}. 

The plan of the paper is as follows. In the next Section we discuss
some details of the calculation. Section 3 presents
our results whereas in the last one we outline some brief conclusions and 
possible prospects.

\section*{2. Parameters}

As for the details of the computation techniques of the relevant Feynman 
amplitudes\footnote{Note that
the analytical expression for the amplitude of the signal process 
(\ref{proc}) is identical to that given in the Appendix
of Ref.~\cite{ioeKosuke} for the hadronic
cases, provided one replaces there $p_U$ and $p_D$ (i.e., the light
quark four-momenta) with $p_{e}$ and $p_{\nu_e}$ (i.e., the electron/neutrino
ones).}, for the choice of structure functions 
as well as for the numerical values of the SM parameters used in this 
paper, we refer the reader to Ref.~\cite{epSM}\footnote{Note that, while
in Ref.~\cite{epSM} several different choices of the 
renormalisation/factorisation scale were adopted, in the present analysis
we stick to the unique `running value' $\mu=\sqrt{\mathaccent94{s}}$, 
that is the CM energy at parton level.}. The only exception is
the top width, which has been modified in order to allow for SUSY decays of
the top quark. Concerning the MSSM parameters,
we assume a universal
soft Supersymmetry-breaking mass \cite{corrMH0iMSSM,corrMHMSSM}
$m_{\tilde u}^2=m_{\tilde d}^2=m_{\tilde q}^2$
and negligible mixing in the stop and sbottom mass matrices,
$A_t=A_b=\mu=0$.
Under these conditions, the one-loop corrections to the masses of the 
MSSM  neutral ${\cal {CP}}$-even
Higgs bosons and to the mixing angle $\alpha$
are introduced via simple relations 
(see Refs.~\cite{0pmLEPLHCSSC,corrMH0iMSSM}), which we have already recalled 
in Ref.~\cite{ioeKosuke}.
For the MSSM charged Higgs mass we have maintained the tree-level relation
mentioned in the Introduction, since one-loop corrections are small compared
to those for the neutral Higgses \cite{corrMHMSSM}.
As it is impractical to cover all possible regions of the
MSSM parameter space $(\MA,\tb)$, we have decided to concentrate
here on the two representative 
values $\tan\beta=1.5$ and 30. and on masses of the pseudoscalar Higgs
boson $A$ in the range 60 GeV $\Ord\MA\Ord$ 220 GeV.

As was done in Ref.~\cite{epSM}, we consider (as an illustration,
see the discussion there) the case
of positron beams from LEP (i.e., of $e^+b$-fusion). The
total CM energy $\sqrt s_{ep}$ of the colliding particles will span 
in the range between 300 GeV
(i.e., around the HERA value) and 2 TeV. However, we will focus our attention
mainly to the case of a possible LEP2$\oplus$LHC accelerator, as illustrated
in the previous Section.

Before proceeding with the discussion of the results, we present in
Tab.~I the cross sections of the signal process
(\ref{bH}) evaluated at the LEP2$\oplus$LHC energy for twenty-four
different sets of structure functions. This is done in order to 
estimated the theoretical error due to the $b$ Parton Distribution
Function (PDF) (see discussion in Ref.~\cite{epSM}). By using the very last 
generations of structure functions to be found in the literature, we estimated
the PDF dependence to be approximately 25\%, with the maximum value 
of the total cross section
differing from the minimum one by 180 fb. We believe such 
uncertainty to be already at the present time a quite small error, so
to motivate further and more detailed simulations (including
hadronisation, detector effects, reducible background \cite{LHC_top,aroma})
of charged Higgs phenomenology at future $ep$ colliders. 

\section*{3. Results}

We present the total cross section rates for process
(\ref{bH}) in Figs.~2--3\footnote{Incidentally, we mention that, as a check
of our results, we have verified that the rates presented here for 
the single-top
subprocess in Narrow Width Approximation (NWA) reproduce quite well
the cross section for on-shell top production
\be\label{on_shell}
e^+ b\ar \bar\nu_e t,
\ee
and are compatible with those presented in Ref.~\cite{epSM} for the
SM decays $t\ar bW^\pm$ once the MSSM top width is consistently
adopted.}. Generally the cross section at
$\tan\beta=30.$ (Fig.~2a) 
is greater than that at $\tan\beta=1.5$ (Fig.~2b). This
is because  the contribution
to the total cross section of subprocess (\ref{VSfusion}) 
is negligible at $\tba$ (Fig.~3a) whereas it is sizable at $\tbb$ (Fig.~3b) 
and also because the top BR into $bH^\pm$ pairs is 
higher at larger $\tan\beta$'s (see Fig.~4). These
features can be interpreted in terms of scalar-fermion vertices.

In the case of diagram 1 in Fig.~1 
the enhancement due to the $\sim m_b\tb$ scalar coupling to the $b$
is greater than the suppression due to the reduced strength $\sim1/\tb$ 
of that to the top quark. In case of diagram 2 in Fig.~1, one should 
recall that, with increasing $\tb$,
the $Ab\bar b$ vertex grows (rather quickly, as $\sim\tb$) and so 
do the $Hb\bar b$ and $hb\bar b$ ones, though less sharply (they
are proportional to $\ca/\cb$ and $\sa/\cb$, respectively). As for the
$W^\pm\Phi H^\pm$ vertex, things go the opposite(same) way for the 
heavy(light) Higgs scalar $H(h)$: that is, the
vector-(neutral) scalar-(charged) scalar coupling 
tends to decrease(increase) with increasing $\tb$, though only for
values of $\MA\Ord140$ GeV. For heavier $\MA$'s the role of the
two scalars is interchanged. Finally, the $W^\pm A H^\pm$ interaction shows no
dependence on the MSSM parameters.

Fig.~3  emphasises the point that, at
LEP$\oplus$LHC energies and for 
a yearly luminosity of 1 fb$^{-1}$, 
a charged Higgs particle of less 
than the top mass (corresponding to $M_A\approx140$ GeV)
can be largely produced.
For $\MA\OOrd140$ GeV, the rate begins to be 
very small with a strong decrease of the 
cross sections for an increasing Higgs mass, this being 
mostly due to the fact that the dominant
single-top diagram gets small because of the suppressed BR of top quarks
into $bH^\pm$ pairs. Where the vector-scalar fusion diagram plays a  
dominant role (i.e., for $\MA\OOrd150$ GeV), are 
phase space effects that are more relevant as compared to those
due to the couplings (the latter have a complicate dependence on $\MA$
for different vertices and scalar bosons as well), since the overall
feature is a decrease of the cross section at larger $\MA$'s. 

The dependence of the production rates on the CM energy
$\sqrt s_{ep}$ is governed by the kinematic suppression on the single-top
production and at low energies, say $\sqrt s_{ep}\Ord 500$ GeV, 
the cross section falls to
negligible scales for all combinations in the plane $(\MA,\tb)$. 
In general, although the rate of process
(\ref{bH}) is small at existing collider energies ($\sqrt{s}_{ep}\approx
300$ GeV at DESY leads to a total cross section of less than 0.1 fb, which
is negligible given the current integrated luminosity of about 20
pb$^{-1}$ \cite{desy} at each of the two experiments), it increases
markedly near the TeV scale. At the LEP2$\oplus$LHC energy it is
easily observable already after one year of running
as long as single-top production dominates.
As a matter of fact, when this is no longer the case (i.e., 
when $\MHpm\OOrd m_t$, the `critical' heavy range), production rates
fall below detection level. In fact, the cross sections never exceed the
1 fb value. This makes immediately the
point that the discovery potential of future $ep$ colliders is
realistically confined to the intermediate $\MHpm$ range only, where
the coverage furnished by the LHC and the NLC will probably 
be more than adequate.

However, the production mechanism is here different, as it also proceeds 
(other than via top decays) through additional diagrams involving the neutral
Higgses whose effects are perceptible over a sizable portion of the MSSM
parameter space, provided $\tb$ is large enough (see Fig.~2d). This is 
particularly
true at small values of $\sqrt s_{ep}$ (where the kinematic threshold 
suppression on 
$t\ar bH^\pm$ is active, i.e., $\sqrt{\mathaccent94{s}}\sim m_t$)
and $\MA$ as well. At those energies though, the total cross section of process
(\ref{bH}) is too low. In contrast, this is no longer the case at 
LEP2$\oplus$LHC energies, where the effects of graph 2 are still significant
(for large $\tb$'s) 
and act on a comfortably large total cross section. On its own, subprocess
(\ref{VSfusion}) yields (at $\sqrt s_{ep}\approx 1.7$ TeV) a rate
of approximately 2 fb (for $\MA=140$ GeV and $\tbb$). 
However, a somewhat stronger effect 
appears through the (negative) 
interference between the two graphs in Fig.~1, reducing
the single-top rates by $-10\%$ or so. This is presumably the effect that
one should search for, even because it will probably not be
possible to separate efficiently the two components 
(\ref{tbH})--(\ref{VSfusion}) of the cross section, as
the charged Higgses produced would decay leptonically
with the neutrinos escaping the detectors. Therefore, 
the $t\ar bH^\pm$ resonance 
cannot be reconstructed and exploited to remove single-top events from
the complete sample.
For 1 fb$^{-1}$ of yearly luminosity, the above rates mean that some 
5 events out of
the 59 expected from single-top production should be missing\footnote{Note 
that, if the electron(positron) beam will have
a 50 GeV energy, this
yielding $\sqrt s_{ep}\approx$ 1.2 TeV, so to increase the luminosity 
(see Ref.~\cite{LHC_phys}) by a factor of ten, 
one would then rely on a statistically
more significative sample, as at that energy the depletion due to 
interference effects is around 8\% (see Figs.~2d). 
However, we do not consider here such a possibility.}. 
Furthermore, one should recall that these `production' rates 
can be fully exploited 
also at `decay' level as the $\tau\nu_\tau$ channel 
has a BR of practically one at large $\tb$'s.
Such effect could well be used to test possible anomalous
couplings in the Higgs sector of the MSSM 
and/or in constraining possible gauge violations
affecting the $W^\pm\Phi H^\pm$ vertex.

In the remainder of the paper, since the only $H^\pm$ mass range that can be
explored
at future TeV $ep$ colliders is below the value of $m_t$, we will consider
$\tau\nu_\tau$ decays of the charged Higgs boson only: that is, the two-to-four
body reaction
\be\label{leptonic}
e^+ b\ar \bar\nu_e bH^{+}\ar \bar\nu_e b \nu_\tau\tau^+.
\ee
The signature that one should expect from this process would then be
a $\tau$-jet (which we assume easily distinguishable from those originated
by quarks and gluons), a $b$-jet (which we assume to be vertex tagged with
efficiency close to unity) and appreciable missing momentum, and the signal
should be revealed as a clear excess with respect to the rates due to 
SM processes (the recalled lepton universality breaking signal).

Fig.~5 plots the differential distributions in various
kinematic quantities which can be reconstructed from the detectable
particles in the final state of process (\ref{leptonic}). 
The distribution in transverse momenta $p_T$ shows that neither cuts in
$p_T$ nor cuts in $p_T^{miss}$ will affect the total cross section
dramatically, whereas that of $\Delta R$, the azimuthal-pseudorapidity
separation defined by $\Delta R = \sqrt{(\Delta\phi)^2+(\Delta\eta)^2}$
(where $\phi$ is the azimuthal angle and $\eta$ the pseudorapidity)
indicates that the requirement of an isolated lepton may strongly affect
the event rate.
The majority of events are found within $\Delta R\Ord 1.5$, which is about
90 degrees in the azimuthal angle. This is because the bottom quark jet
and the tau come from the energetic top quark. Thus, at lower energies the
azimuthal-pseudorapidity spread in the top quark decay products will be
larger and hence the requirement of an isolated lepton not so severe.
The distribution of the missing transverse momentum is small at low
missing $p_T$ and indicates that the charged current cut in missing
transverse momentum will not affect the event rate significantly.

Tab.~II shows the total cross section after the
acceptance cuts. The following constraints were implemented (see
\cite{LHC_top} for discussions): $p_T^{\tau^+}, p_T^{b}>20$ GeV,
$p_T^{miss}>10$ GeV and $\Delta R_{\tau^+,b}>0.7$. We have not implemented
any cuts on the pseudorapidity, as the events are all concentrated in the
detectable $|\eta|$ region: see the spectra
in the two lower frames of Fig.~6.

In Fig.~6 we also plot the distributions in 
the invariant mass of the only visible pair of particle momenta, the $b$ and
$\tau$-ones, i.e., $M_{b\tau}$. 
This is done in order to possibly aid further the signal 
selection, as this cannot rely on the kinematic reconstruction of the
charged Higgs boson mass, because of the $\tau$-neutrino.
In particular we would like to point out that there is a kinematic interplay
between, on the one hand, the top, tau and bottom masses and, on the other
hand, that of the boson produced in the top decay, inducing a cut-off on the
maximum value of $M_{b\tau}$. This should clearly be different for the
ordinary SM-like backgrounds, particularly that due to single-top production
followed by $t\ar bW^\pm$ (the dominant one, see Ref.~\cite{epSM}). In 
general, assuming that both the top quark and the decay
boson are on-shell, the cut-off is given by $M_{b\tau}^{max} =
\sqrt{m_t^2+m_b^2+m_\tau^2-M_{V}^2}$, with $M_V=\MHpm$ or $M_{W^\pm}$.
For example, at low $M_A$, the cut-off in the roughly
triangular distributions in $M_{b\tau}$ is close to the top mass as
expected, whereas at high $M_A$, as the charged Higgs mass tends to the
top mass, the cut-off is smaller since the momenta carried by the bottom
quark become less energetic. For comparison, 
in Fig.~7 we present the same $M_{b\tau}$
distribution for the background processes 
\be\label{W}
e^+ b\ar \bar\nu_e b \tau^+ \nu_\tau,
\ee
and
\be\label{W_CC}
e^+ \bar b\ar \bar\nu_e \bar b \tau^+ \nu_\tau,
\ee
both proceeding via a $W^{+(*)}\ar \tau^+ \nu_\tau$ splitting (see
Figs.~1c and d of Ref.~\cite{epSM}, respectively).
To facilitate the comparison between the two figures, the
normalisation has been set to unity. We see that the spectrum
of the MSSM signal is significantly harder(softer) that the SM-like one
from process (\ref{W}) for smaller(larger) values of $\MA$, at all
$\tan\beta$. There is a sort of degeneracy between the two processes
(\ref{leptonic}) and (\ref{W}) for $\MA=100$ GeV, rather than for
smaller $\MA$'s (thus for $\MHpm$'s closer to $M_{W^\pm}$). This is due to
the additional diagrams entering in the latter reaction, which do not
suffer from the kinematic cut-off. Also note that for the background there is
no dependence of the shape on the actual values of the MSSM parameters.
As for events of the type (\ref{W_CC}),
the distribution is rather flat, with no evident kinematic peak.
Thus, apart for $\MA\approx100$ GeV,
the $M_{b\tau}$ spectrum should indeed help in disentangling the 
$H^\pm$ signals from the irreducible background.

The total cross sections of process (\ref{W}) are 
presented in Tab.~III, for the same choice of the $(\MA,\tb)$ parameters
as in the previous one. We do not reproduce here the rates for process
(\ref{W_CC}) as these are two orders of 
magnitude smaller, around 4.8 fb, and with no dependence on
$\MA$ and/or $\tb$,  confirming that the background that does not
involve the on-shell top production will not affect the detection of
$H^\pm$ signals at all, even in case of poor performances in measuring
the jet charge of the $b$-jet.
The dependence of the background (\ref{W}) on the MSSM parameters can be
traced back to the simple fact that the greater the BR for the
top decay into $H^\pm$ the smaller the one for the
background process $t\rightarrow bW^+$. 
The dependence entering in the total cross section of process (\ref{W})
through the neutral Higgs mediated diagrams (see Fig.~1c of Ref.~\cite{epSM})
is indeed negligible.

It can be seen that for
$\tan\beta=1.5$ the lepton universality breaking signal is significant
over the background for $M_A$ up to about 100 GeV, whereas at
$\tan\beta=30.$ the signal is significant up to 120 GeV.
Therefore, a combination of event rate counting and $M_{b\tau}$ distribution
studies should allow for the detection of charged Higgs bosons of the MSSM
over a large portion of the $(\MA,\tb)$ plane.

\section*{4. Summary and conclusions}

The phenomenology of charged MSSM Higgs bosons $H^\pm$ produced from
initial state bottom sea quarks at future $ep$ colliders was studied,
mainly focusing our attention to the case of the planned
LEP2$\oplus$LHC accelerator with the
positron(electron) beam energy of 100 GeV and the proton one of 7
TeV. The design of such collider was already 
proposed in 1990 as a possible extension
of the LEP and LHC programmes at CERN and the physics of such machines
has been the object of a recent renewed interest. The motivation for our
analysis was twofold. First, it is quite clear that neither the LHC nor
the NLC will probably be able to cover adequately the heavy mass range
of the $H^\pm$ scalar (when $\MHpm\OOrd m_t$). Second, in the accessible
intermediate mass range (i.e., $\MHpm\Ord m_t$), charged Higgs bosons
are produced in either top decays (LHC) or via $\gamma\ar H^+H^-$ splitting
(NLC), so that the insight on the phenomenology
of the 
Higgs sector of the MSSM is only confined to a test of the $\beta$-dependence
entering in the $H^\pm bt$ vertex, as the $\gamma H^+H^-$ one only depends
on the electromagnetic coupling constant. In our opinion,  
it was therefore important to assess the potential of future 
$ep$ colliders in the above respects. 

Assuming 1 fb$^{-1}$ of integrated luminosity per year of running,
the detectable rate was found to vary in the range 2--300 events per annum,
depending on the Higgs masses and $\tan\beta$. The uncertainty due to the
structure functions was found to be rather small already at the present time,
around 25\%, and is expected to diminish significantly before new $ep$
machines will enter operation. The majority of the Higgs boson
production, in the range of collider energies relevant to our study
(i.e., between 300 and 2000 GeV), occurs
for values of $\MHpm$ below the top mass, where the dominant
decay mode of charged Higgs bosons is into $\tau\nu_\tau$ pairs. For such
values of $\MHpm$ the event rate is primarily due to
single-top quark production and decay in the charged
current process $e^+b\rightarrow\bar{\nu}_et\ar \bar{\nu}_e b H^+$. 

Although
the production of $H^\pm$ scalars via a top decay is also the signature typical 
at the LHC, the additional feature of 
MSSM charged Higgs boson production at future $ep$ colliders is a sizable 
contribution from vector-scalar fusion diagrams of the type 
$e^+ b\ar \bar\nu_e b W^{\pm *}\Phi^{*}\ar \bar\nu_e bH^{+}$ involving
the neutral Higgs bosons of the MSSM (i.e., $\Phi=H,h,A)$ in virtual state,
provided $\tan\beta$ is large.
This is particularly true for small $\MA$'s and for
values of the CM energy of the $ep$ 
collisions below 500 GeV (when the `partonic' CM energy 
is often below the single-top production threshold), where corrections
to the cross section can amount to up +20\% of the single-top rate. However,
at typical LEP2$\oplus$LHC energies the effect is still significant
and should be visible through
a negative interference between single-top and vector-scalar fusion diagrams,
which depletes the single-top event rate by up to $-10\%$ or so,
for $140~{\mbox{GeV}}\Ord\MA\Ord160$ GeV. Such effect clearly
represents a test of the Higgs sector of the MSSM as well as a constraint
on possible gauge violation effects which could manifest themselves in the
$W^\pm\Phi H^\pm$ vertex.
For values of $\MHpm$ in the heavy range, the single-top production rates
fall at negligible levels and the kinematic suppression on the vector-fusion
mechanism is such that the production cross section is below detection level,
even for optimistic luminosities. Therefore, as for heavy charged Higgses of
the MSSM, no further $H^\pm$ detection potential other than that already
provided by the LHC and 
NLC should be expected by future $ep$ colliders at the TeV scale.

Finally, several kinematical quantities associated with the momenta which
can be reconstructed from the visible 
particles of the signature $b\tau^+ E_{miss}$ were studied.
In general, LHC-type detectors should provide an adequate coverage (in 
pseudorapidity and transverse momentum) of the phase space available to 
events of the type 
$e^+b\ar \bar\nu_e b H^+\ar \bar\nu_e b\tau^+\nu_\tau$. In addition, the
invariant mass of the $b\tau^+$-pair can be profitably exploited in separating 
the $H^\pm$ signal events from the irreducible SM-like electroweak background,
which we have evaluated consistently. Indeed, the 
lepton universality breaking signature can be used as a mean to recognise MSSM
charged Higgs events for $A$ masses up to 100--120 GeV (that is 
$\MHpm\approx130-145$ GeV) for both small and, more markedly, large values
of $\tan\beta$.

We believe that charged Higgs boson phenomenology in the context of MSSM
can be a relevant experimental issue at future $ep$ colliders, and we look
forward to additional studies of (possibly) new production mechanisms as well as
more detailed simulations, including detector and hadronisation effects.
To this end, the {\tt FORTRAN} programs used for this analysis can be obtained from
the authors upon request.

\section*{5. Acknowledgements}

We thank Lorenzo Diaz-Cruz for useful discussions.
SM is grateful to the UK PPARC and KO to Trinity College and the Committee
of Vice-Chancellors and Principals of the Universities of the United
Kingdom for financial support. SM Also thank the Theoretical
Physics Department in Lund (Sweden), where part of this work was carried
out under a grant of the Italian Institute of Culture `C.M. Lerici' 
(Ref. \#: Prot. I/B1 690).

\goodbreak

\vfill
\newpage

\subsection*{Table Captions}
\begin{description}

\item{[I] } Total cross sections for process
(\ref{bH}) at LEP2$\oplus$LHC energies for twenty-four different sets of
structure functions.  Errors are
as given by VEGAS (the same statistics points
were used for the {\tt NCALL} and
{\tt ITMX} parameters) \cite{VEGAS}. As representative values of the
MSSM parameters we have used $M_A=60$ GeV and $\tan\beta=1.5$.

\item{[II] } Total cross section for process (\ref{leptonic})
at the LEP2$\oplus$LHC collider, for a selection of Higgs masses. The
structure function set MRS(A) was used. Errors are as given by VEGAS
\cite{VEGAS}. The following acceptance cuts were implemented:
(i) $p_T^{\tau^+}, p_T^{b}>20$ GeV, $p_T^{miss}>10$ GeV
and $\Delta R_{\tau^+,b}>0.7$.

\item{[III] } Total cross section for process (\ref{W})
at the LEP2$\oplus$LHC collider, for a selection of Higgs masses. The
structure function set MRS(A) was used. Errors are as given by VEGAS
\cite{VEGAS}. The following acceptance cuts were implemented:
(i) $p_T^{\tau^+}, p_T^{b}>20$ GeV, $p_T^{miss}>10$ GeV
and $\Delta R_{\tau^+,b}>0.7$.

\end{description}

\vfill
\newpage

\subsection*{Figure Captions}

\begin{description}

\item{[1] } Lowest order Feynman diagrams describing processes (\ref{proc}).
The package MadGraph \cite{tim} was used to produce the PostScript codes. 
Graph 1 refers to single-top production and decay whereas graph 2
corresponds to the vector-scalar fusion mechanism.

\item{[2] } The total cross section $\sigma$ for
process (\ref{bH}) (a,b) and the ratio $R_\sigma$ between this and that of
process (\ref{tbH}) (c,d), with
300 GeV $\leq\sqrt{s}_{ep}\leq$ 2 TeV, 
for $\tan\beta=1.5$ (a,c) and 30. (b,d) and for 
five different values of the pseudoscalar Higgs mass: 
$M_A=60$  GeV (continuous lines), 
$M_A=100$ GeV (short-dashed lines),
$M_A=140$ GeV (dotted lines),
$M_A=180$ GeV (dot-dashed lines) and  
$M_A=220$ GeV (long-dashed lines).
The structure function set MRS(A) was used.

\item{[3] } The total cross section for
process (\ref{leptonic}) at the LEP2$\oplus$LHC CM energy as a function of the
pseudoscalar Higgs mass in the range 60 GeV $\leq M_A\leq$ 220 GeV, for
$\tan\beta=1.5$ (continuous line) and
$\tan\beta=30.$ (dashed line). The structure function set MRS(A) was used.

\item{[4] } The br\-an\-ch\-ing ra\-tios of the top quark and charged
Higgs boson
as a function of the charged(pseudoscalar) Higgs mass in the range
$100(60)~\mbox{GeV}\Ord M_{H^\pm}(M_A)\Ord252(240)$ GeV, for $\tba$ and 30.

\item{[5] } Differential distributions 
for process (\ref{leptonic}) at the LEP2$\oplus$LHC CM energy 
in the following variables (clockwise):
1.~$\Delta R_{b\tau}$, the azimuthal-pseudorapidity
separation of the $b\tau$-pair;
2.~$p_{T,miss}$, the missing transverse momentum;
3.~$p_{T,\tau}$, the transverse momentum of the $\tau$-lepton;
4.~$p_{T,b}$, the transverse momentum of the $b$-quark;
for $\tan\beta=1.5$ (a) and 30. (b) and for 
three different values of the pseudoscalar Higgs mass:
$M_A=60$ GeV (continuous lines), 
$M_A=100$ GeV (dashed lines) and 
$M_A=140$ GeV (dotted lines). 
The normalisation is to unity. The structure function set MRS(A) was used.

\item{[6] } Differential distributions 
for process (\ref{leptonic}) at the LEP2$\oplus$LHC CM energy 
in the following variables (from top to bottom):
1.~$M_{b\tau}$,
the invariant mass of the $b\tau$-pair;
2.~$|\eta_b|$, the absolute value of the pseudorapidity of the $b$-quark;
3.~$|\eta_\tau|$, the absolute value of the pseudorapidity of the 
$\tau$-lepton;
for $\tan\beta=1.5$ (a) and 30. (b) and for 
three different values of the pseudoscalar Higgs mass:
$M_A=60$ GeV (continuous lines), 
$M_A=100$ GeV (dashed lines) and 
$M_A=140$ GeV (dotted lines). 
The normalisation is to unity. The structure function set MRS(A) was used.

\item{[7] } Differential distributions in 
the invariant mass of the $b\tau$-pair, $M_{b\tau}$, 
for processes (\ref{W}) ($e^+b$ background) and
(\ref{W_CC}) ($e^+\bar b$ background) at the LEP2$\oplus$LHC CM energy
for $M_A=100$ GeV, $\tan\beta=1.5$ (solid lines) and
                   $\tan\beta=30.$ (dashed lines). 
The normalisation is to unity. The structure function set MRS(A) was used.

\end{description}

\vfill
\newpage

\begin{table}
\begin{center}
\begin{tabular}{|c|c|}
\hline
\multicolumn{2}{|c|}
{\rule[0cm]{0cm}{0cm}
$\sigma(e^+b\ar \bar\nu_e b H^+)$}
\\ \hline
\rule[0cm]{0cm}{0cm}
PDFs & $\sigma_{t}$ (fb) \\ \hline\hline
\rule[0cm]{0cm}{0cm}
MRS(A)   & $752.3\pm2.0$ \\
MRS(A')  & $739.5\pm1.9$ \\       
MRS(G)   & $716.9\pm1.9$ \\
MRS(J)   & $769.2\pm2.1$ \\
MRS(J')  & $828.5\pm2.3$ \\
MRS(R1)  & $701.0\pm1.9$ \\
MRS(R2)  & $757.8\pm2.1$ \\
MRS(R3)  & $716.4\pm1.9$ \\
MRS(R4)  & $766.9\pm2.0$ \\
MRS(105) & $673.9\pm1.8$ \\
MRS(110) & $718.4\pm2.0$ \\
MRS(115) & $712.1\pm1.9$ \\
MRS(120) & $772.9\pm2.1$ \\
MRS(125) & $778.5\pm2.1$ \\
MRS(130) & $791.6\pm2.2$ \\
MRRS(1)  & $804.4\pm2.2$ \\
MRRS(2)  & $805.9\pm2.2$ \\
MRRS(3)  & $803.2\pm2.2$ \\
CTEQ(2M) & $782.5\pm2.1$ \\
CTEQ(2MS)& $758.3\pm2.0$ \\
CTEQ(2MF)& $789.7\pm2.1$ \\
CTEQ(2ML)& $843.8\pm2.3$ \\
CTEQ(3M) & $832.9\pm2.3$ \\
CTEQ(4M) & $815.0\pm2.3$ \\ \hline\hline
\multicolumn{2}{|c|}
{\rule[0cm]{0cm}{0cm}
no acceptance cuts}
\\ \hline

\hline\hline
\multicolumn{2}{|c|}
{\rule[0cm]{0cm}{0cm}
LEP2$\oplus$LHC}
\\ \hline

\multicolumn{2}{c}
{\rule{0cm}{1.0cm}
{\Large Table I}}  \\

\end{tabular}
\end{center}
\end{table}

\vfill
\newpage

\begin{table}
\begin{center}
\begin{tabular}{|c|c|c|}
\hline
\multicolumn{3}{|c|}
{\rule[0cm]{0cm}{0cm}
$\sigma_{tot}$ signal (fb)}
\\ \hline
\rule[0cm]{0cm}{0cm}
$M_{A}$ (GeV) & $\tan\beta=1.5$ & $\tan\beta=30.$ \\ \hline\hline
\rule[0cm]{0cm}{0cm}
$60$   & $194.66\pm0.87 $ & $327.1 \pm1.7$   \\
$80$   & $140.66\pm0.62 $ & $243.9 \pm1.3$   \\
$100$  & $76.95 \pm0.40 $ & $149.12\pm0.80$  \\
$120$  & $24.28 \pm0.14 $ & $62.40 \pm0.36$  \\
$140$  & $2.122 \pm0.010$ & $8.656 \pm0.065$ \\
\hline\hline
\multicolumn{3}{|c|}
{\rule[0cm]{0cm}{0cm}
after acceptance cuts}
\\ \hline

\hline\hline
\multicolumn{3}{|c|}
{\rule[0cm]{0cm}{0cm}
LEP2$\oplus$LHC \qquad\qquad\qquad\qquad\qquad MRS(A)}
\\ \hline

\multicolumn{3}{c}
{\rule{0cm}{1.0cm}
{\Large Table II}}  \\

\end{tabular}
\end{center}
\end{table}

\vfill
\newpage

\begin{table}
\begin{center}
\begin{tabular}{|c|c|c|}
\hline
\multicolumn{3}{|c|}
{\rule[0cm]{0cm}{0cm}
$\sigma_{tot}$ $e^+b$ background (fb)}
\\ \hline
\rule[0cm]{0cm}{0cm}
$M_{A}$ (GeV) & $\tan\beta=1.5$ & $\tan\beta=30.$ \\ \hline\hline
\rule[0cm]{0cm}{0cm}
$60$   & $349.1\pm3.6$ & $341.3\pm3.7$  \\
$80$   & $374.6\pm4.1$ & $358.1\pm4.3$  \\
$100$  & $390.9\pm4.3$ & $384.5\pm4.6$  \\
$120$  & $416.6\pm4.7$ & $413.4\pm5.0$  \\
$140$  & $446.5\pm6.9$ & $439.9\pm6.3$  \\
\hline\hline
\multicolumn{3}{|c|}
{\rule[0cm]{0cm}{0cm}
after acceptance cuts}
\\ \hline

\hline\hline
\multicolumn{3}{|c|}
{\rule[0cm]{0cm}{0cm}
LEP2$\oplus$LHC \qquad\qquad\qquad\qquad\qquad MRS(A)}
\\ \hline

\multicolumn{3}{c}
{\rule{0cm}{1.0cm}
{\Large Table III}}  \\

\end{tabular}
\end{center}
\end{table}

\vfill
\clearpage

\begin{figure}[p]
~\epsfig{file=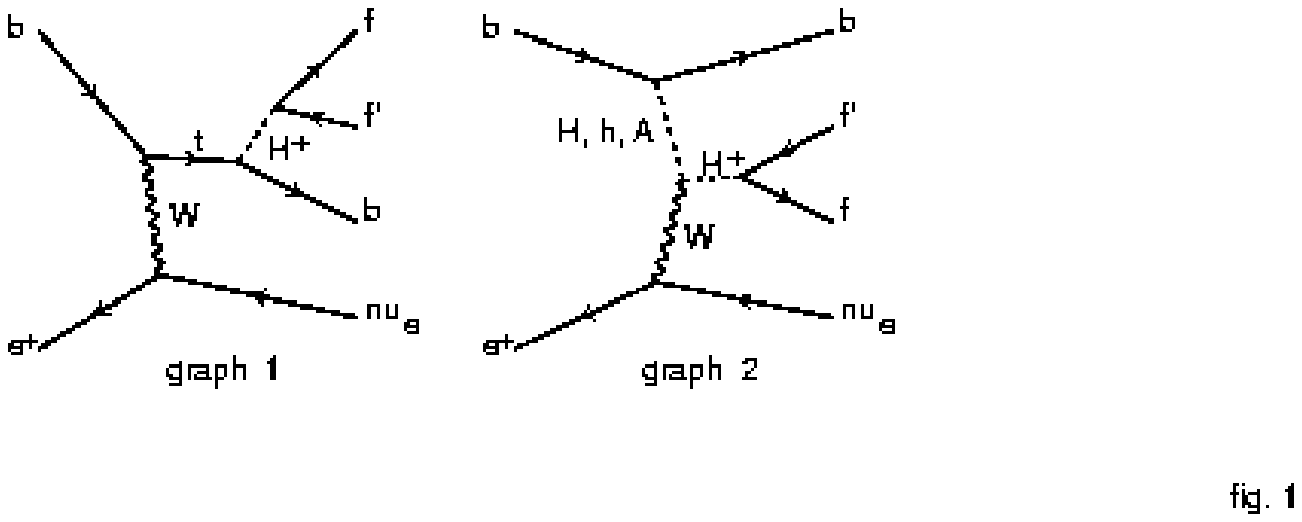,height=23cm}
\vspace*{2cm}
\end{figure}
\vfill
\clearpage

\begin{figure}[p]
~\epsfig{file=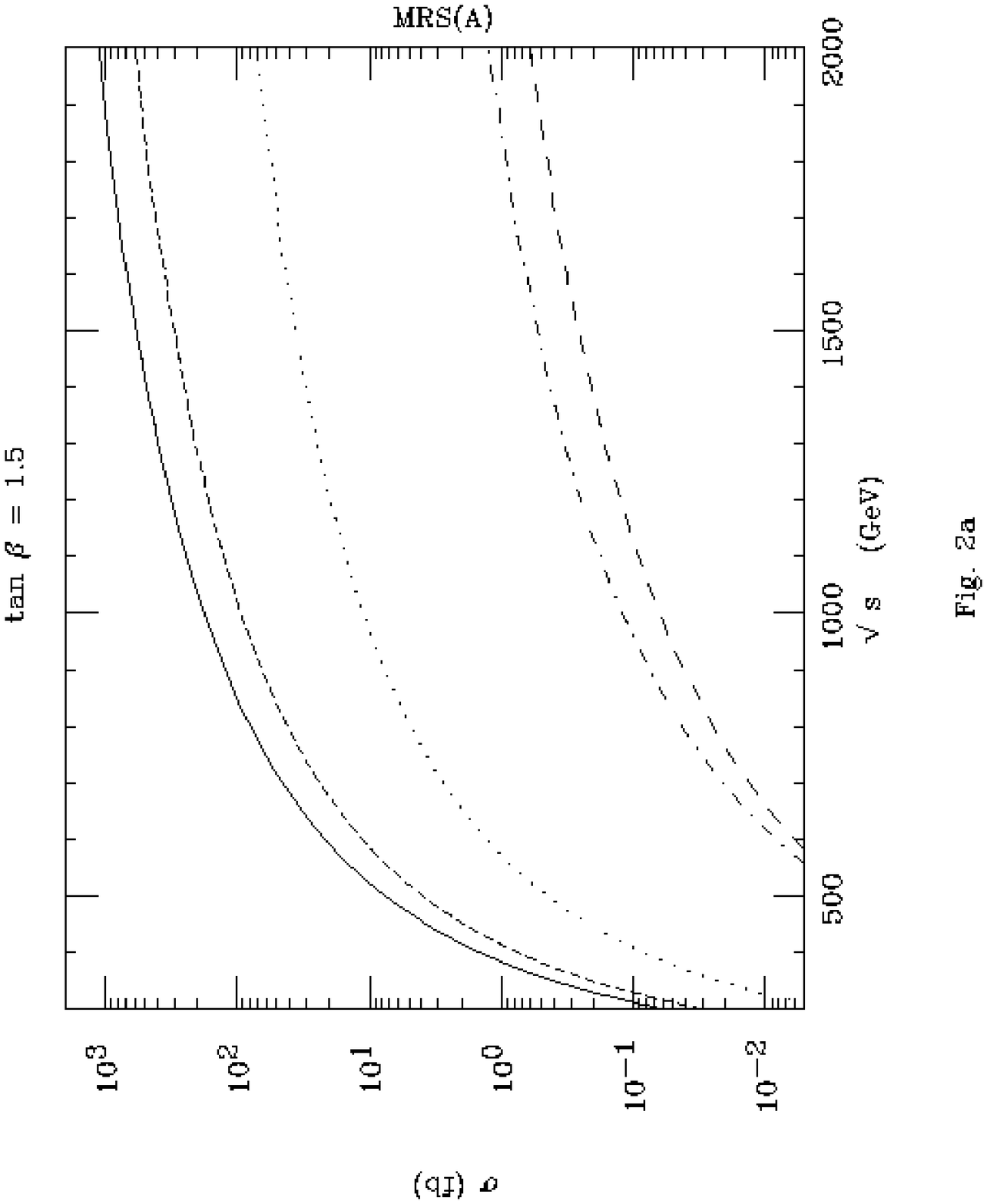,height=23cm,angle=180}
\vspace*{2cm}
\end{figure}
\vfill
\clearpage

\begin{figure}[p]
~\epsfig{file=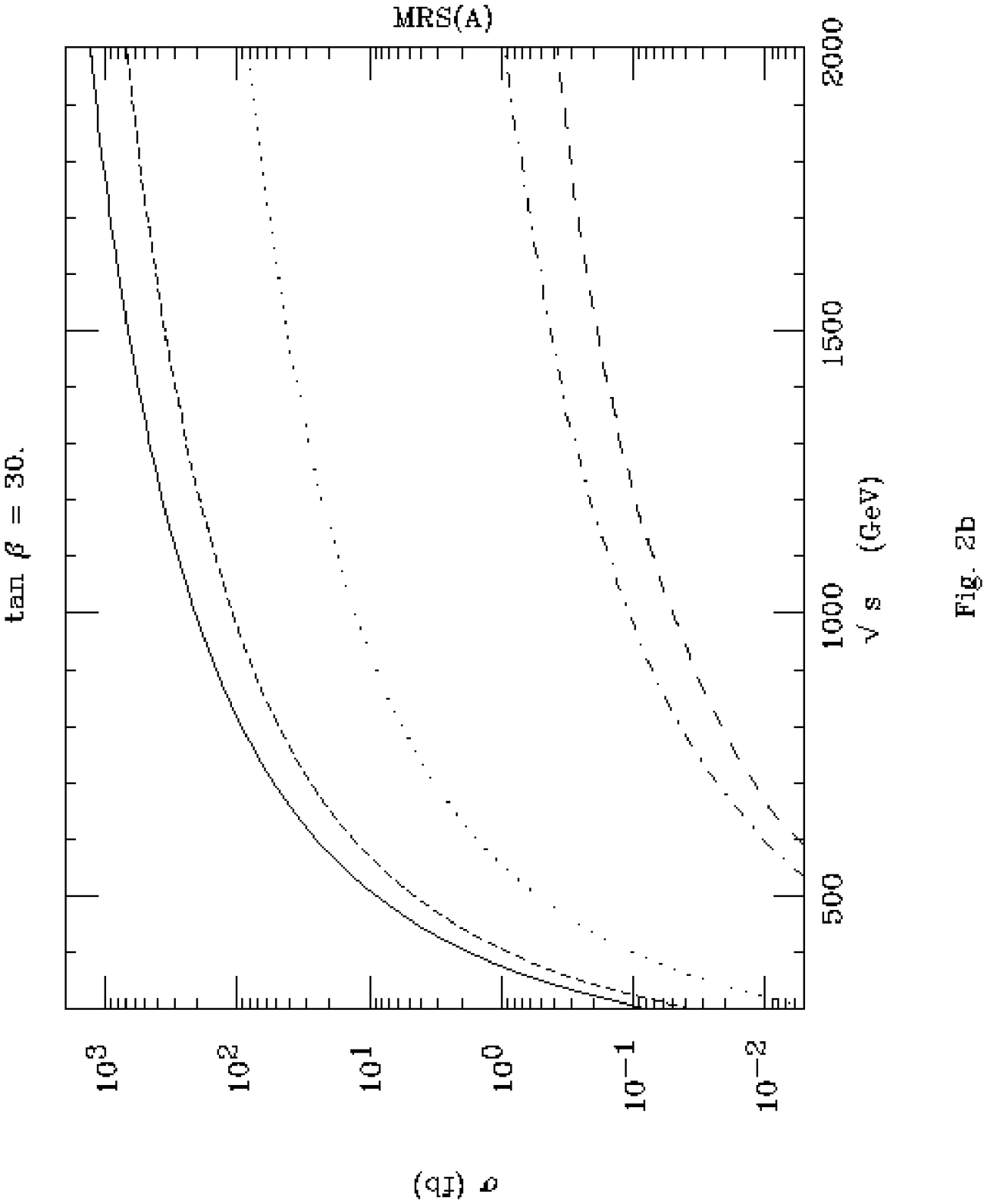,height=23cm,angle=180}
\vspace*{2cm}
\end{figure}
\vfill
\clearpage

\begin{figure}[p]
~\epsfig{file=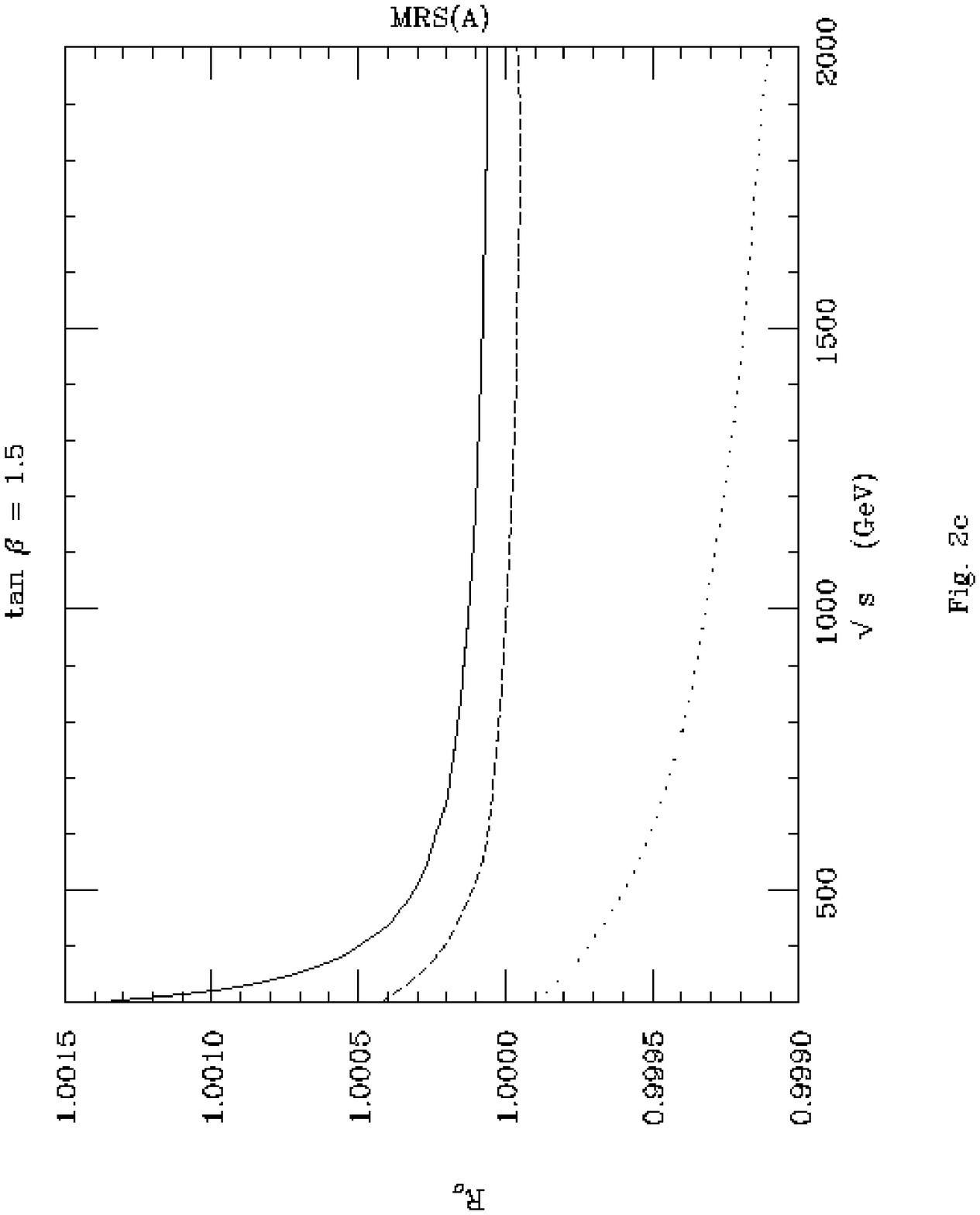,height=23cm,angle=180}
\vspace*{2cm}
\end{figure}
\vfill
\clearpage

\begin{figure}[p]
~\epsfig{file=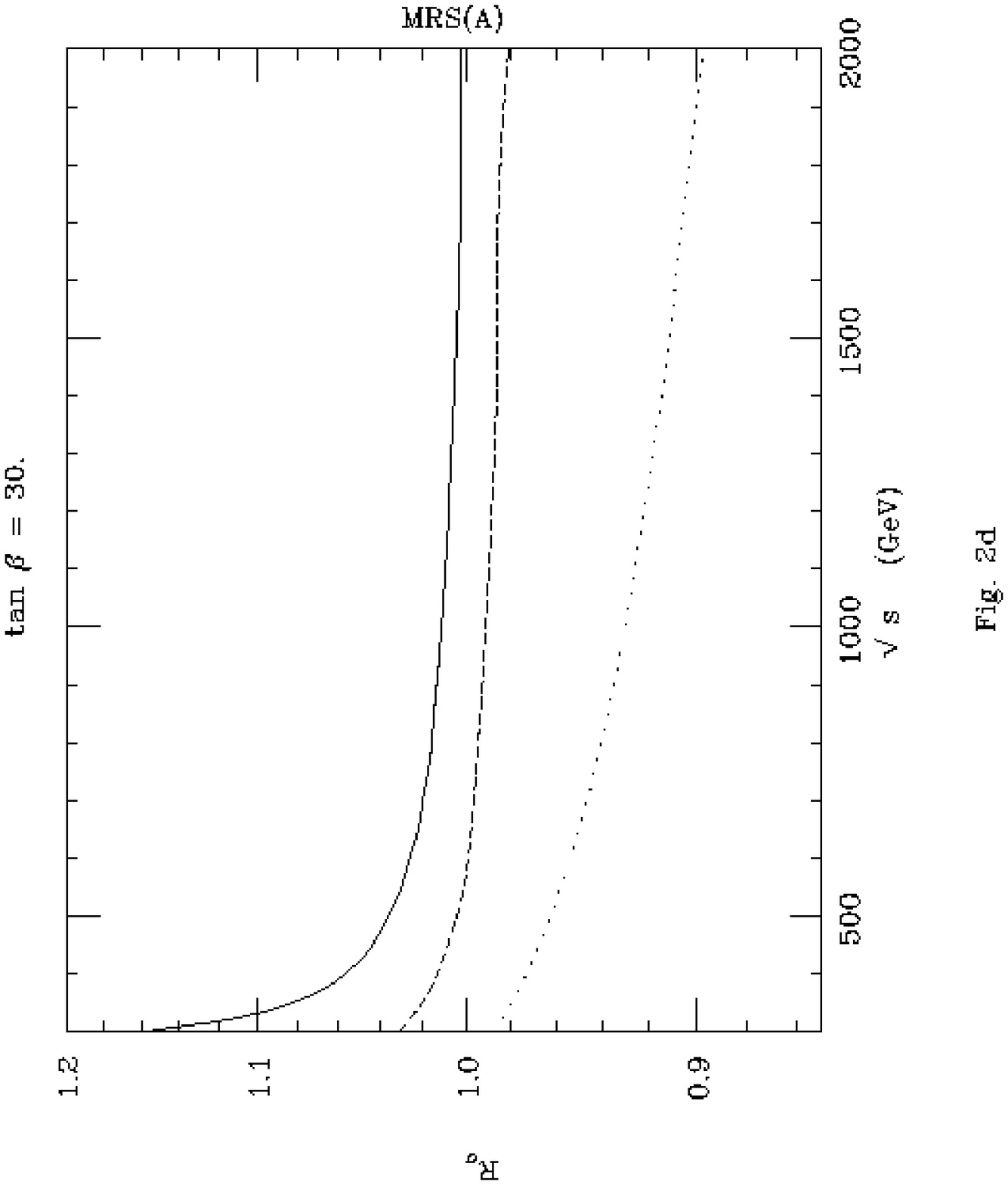,height=23cm,angle=180}
\vspace*{2cm}
\end{figure}
\vfill
\clearpage

\begin{figure}[p]
~\epsfig{file=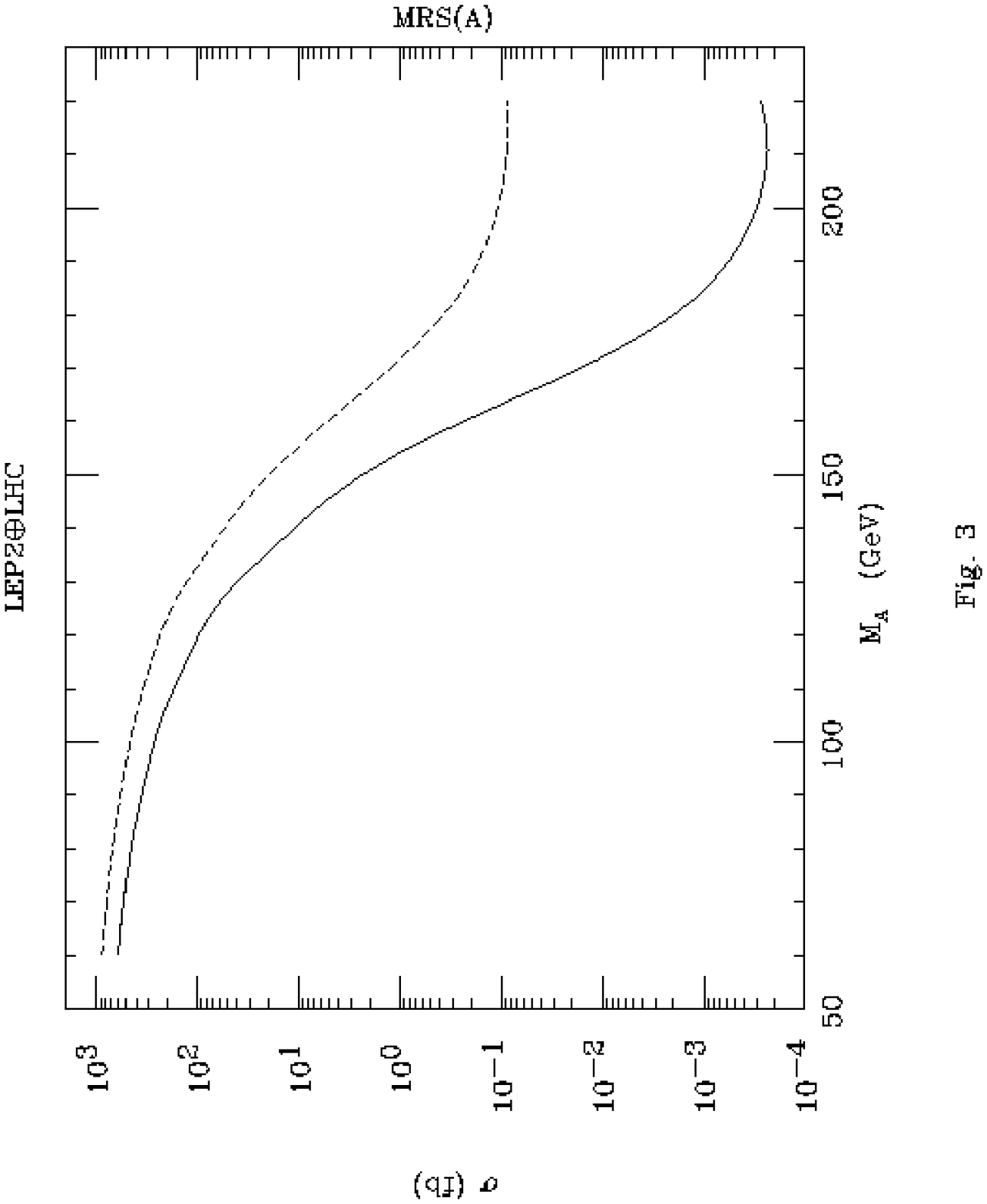,height=23cm,angle=180}
\vspace*{2cm}
\end{figure}
\vfill
\clearpage

\begin{figure}[p]
~\epsfig{file=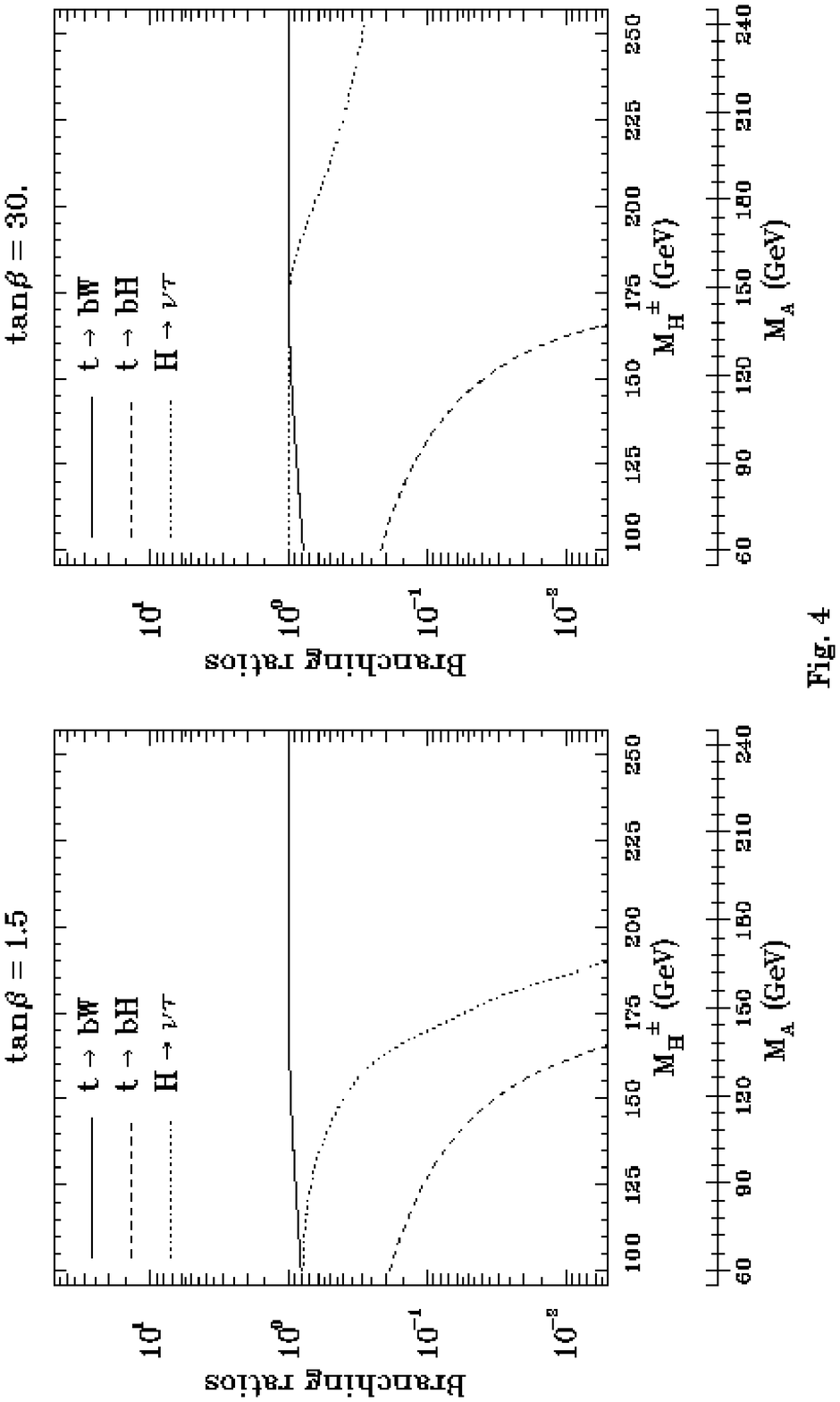,height=23cm}
\vspace*{2cm}
\end{figure}
\vfill
\clearpage

\begin{figure}[p]
~\epsfig{file=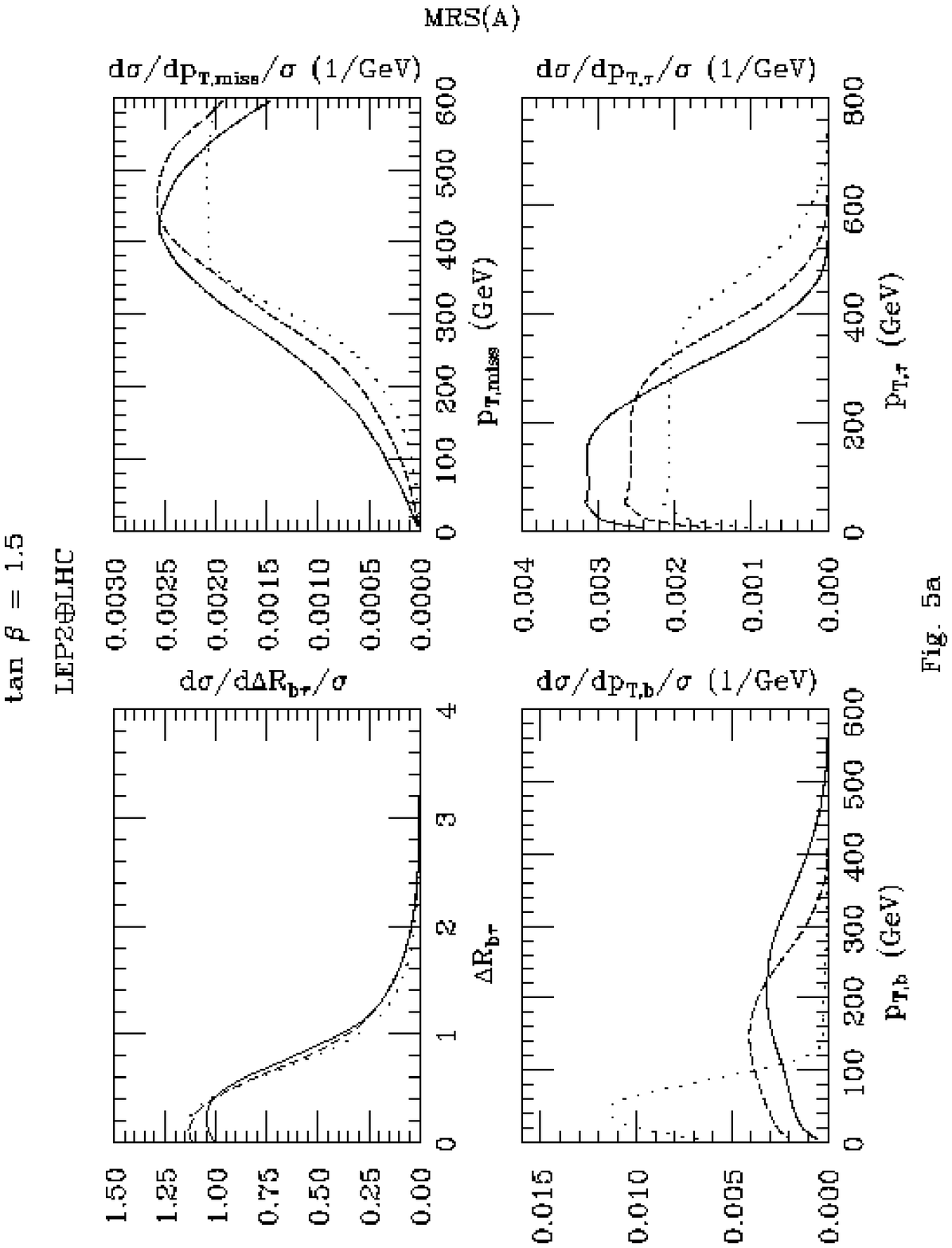,height=23cm,angle=180}
\vspace*{2cm}
\end{figure}
\vfill
\clearpage

\begin{figure}[p]
~\epsfig{file=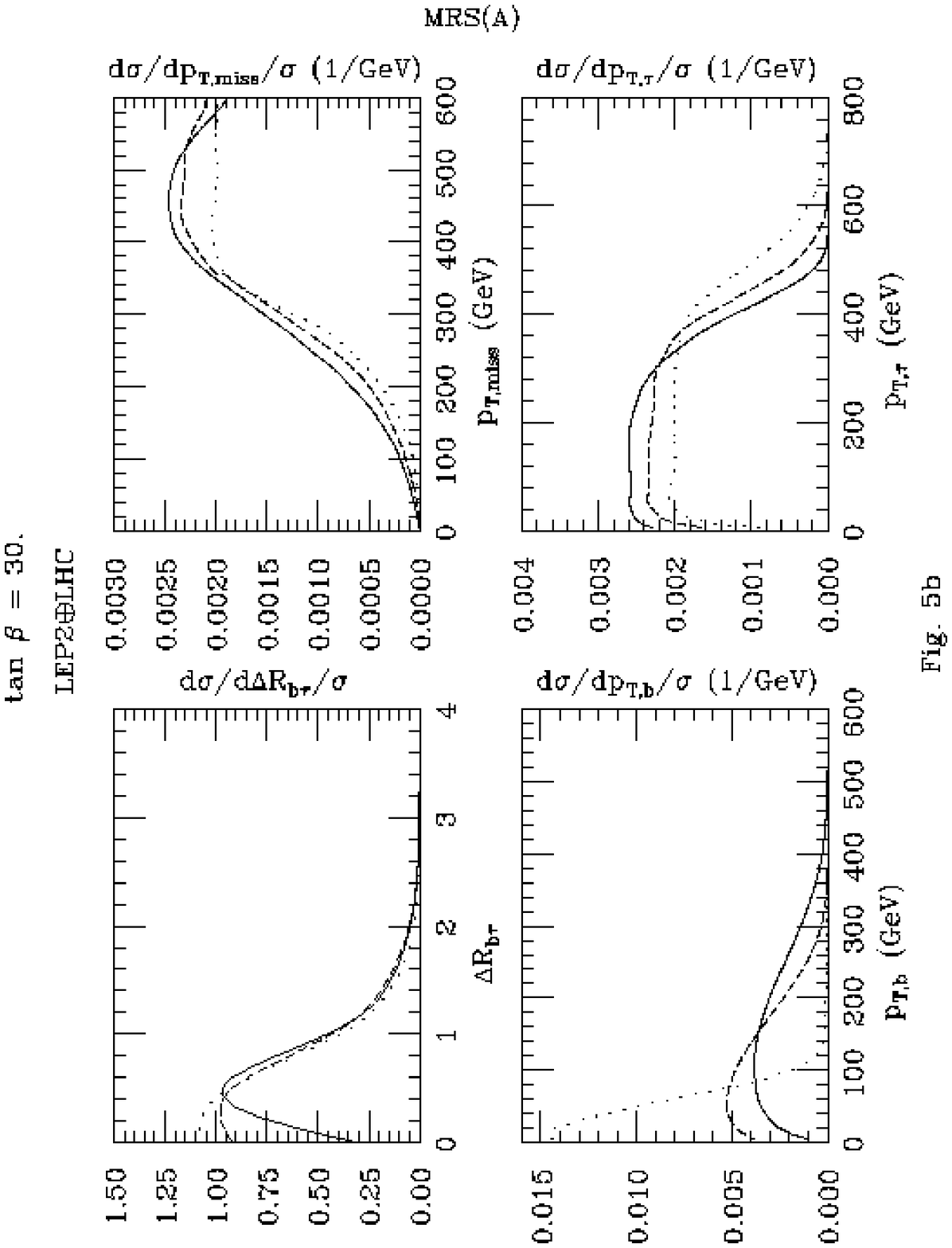,height=23cm,angle=180}
\vspace*{2cm}
\end{figure}
\vfill
\clearpage

\begin{figure}[p]
~\epsfig{file=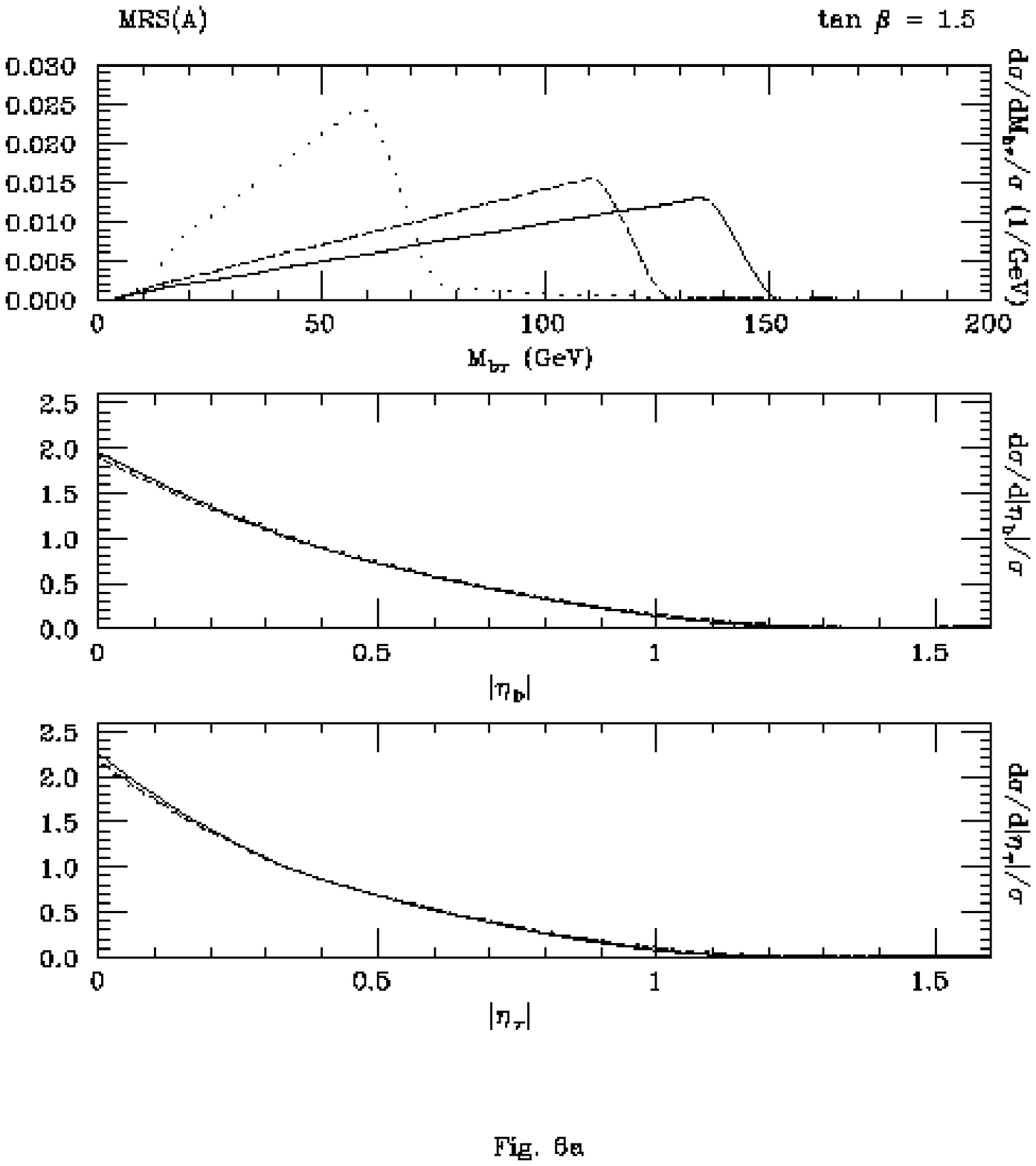,height=23cm}
\vspace*{2cm}
\end{figure}
\vfill
\clearpage

\begin{figure}[p]
~\epsfig{file=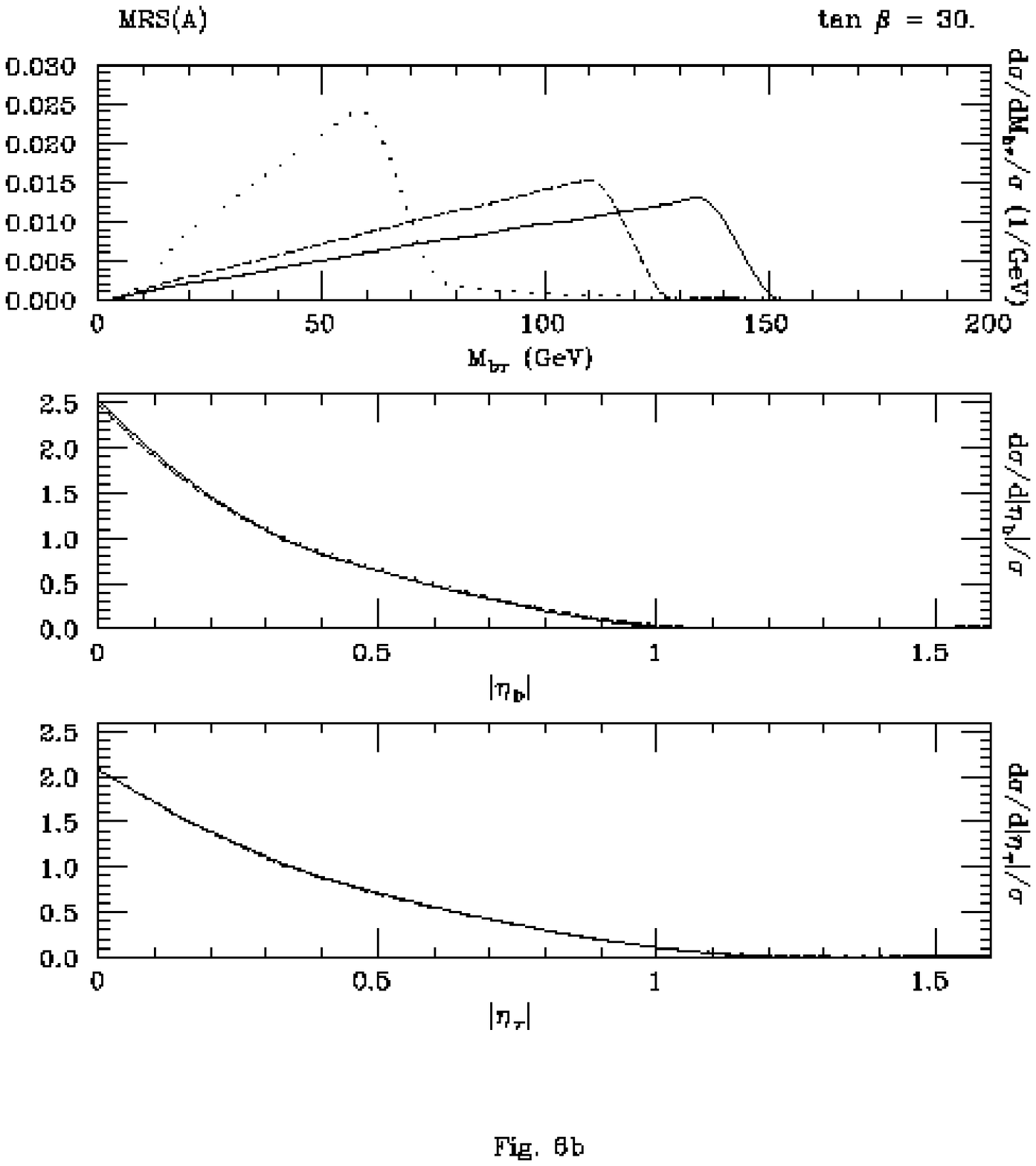,height=23cm}
\vspace*{2cm}
\end{figure}
\vfill
\clearpage

\begin{figure}[p]
~\epsfig{file=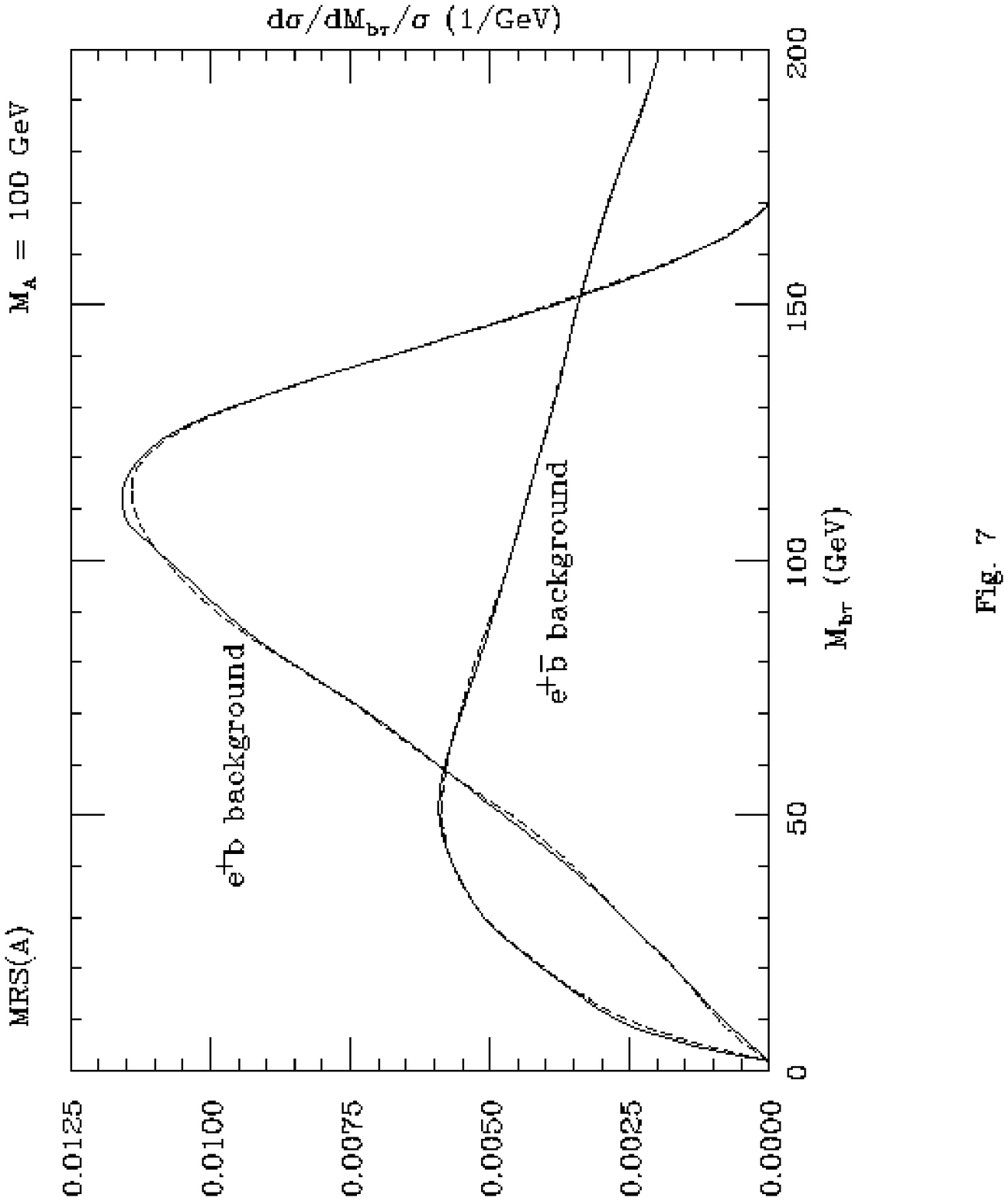,height=23cm,angle=180}
\vspace*{2cm}
\end{figure}

\vfill

\begin{thebibliography}{99}

\bibitem{guide} J.F.~Gunion, H.E.~Haber, G.L.~Kane and S.~Dawson,
                ``The Higgs Hunter Guide''
                (Addison-Wesley, Reading MA, 1990).

\bibitem{Alimit} See, e.g.:\\
The Aleph Collaboration, preprint CERN-PPE/97-071, June 1997.

\bibitem{unitarity} P. Langacker and H.A. Weldon, \prl 52 1984 1377;\\
H.A. Weldon, {\it Phys. Rev.} {\bf D30} (1984) 1547; \pl B146 1984 59.

\bibitem{ioejames} S. Moretti and W.J. Stirling, {\it Phys. Lett.}
{\bf B347} (1995) 291; Erratum, {\it ibidem} {\bf B366} (1996) 451.

\bibitem{wrk41} A. Djouadi, J. Kalinowski and P.M. Zerwas,
                \zp C57 1993 569.

\bibitem{wrk112} S. Komamiya, \pr D38 1988 2158.

\bibitem{eeHH} S. Moretti and K. Odagiri, {\it J. Phys.} {\bf G23} (1997) 537.

\bibitem{lep2w} Proceedings of the Workshop `Physics at LEP2',
eds. G. Altarelli, T. Sj\"{o}strand and F. Zwirner,
CERN Report 96-01.

\bibitem{ATLAS} ATLAS Technical Proposal, CERN/LHC/94-43 LHCC/P2, December 
1994.

\bibitem{CMS} CMS Technical Proposal, CERN/LHC/94-43 LHCC/P1, December 1994.



\bibitem{guide81} J.F. Gunion, H.E. Haber, F.E. Paige, W.-K. Tung and
S.S.D. Willenbrock, \np B294 1987 621.

\bibitem{ioeKosuke} S. Moretti and K. Odagiri, \pr D55 1997 5627.

\bibitem{guide84}  J.F. Gunion, H.E. Haber, S. Komamiya, H. Yamamoto and
A. Barbaro-Galtieri, in Proceedings of the `1987 Berkeley Workshop
on Experiments, Detectors and Experimental Areas for the Supercollider',
eds. R. Donaldson and M. Gilchriese (World Scientific, Singapore,
1988).

\bibitem{guide23} J.F. Gunion and H.E. Haber, \np B272 1986 1.

\bibitem{guide24} J.F. Gunion and H.E. Haber, \np B278 1986 449.

\bibitem{guide132} J.F. Gunion and H.E. Haber, \np B307 1988 445.

\bibitem{ee500} Proceedings of the Workshop
`{$e^+e^-$ Collisions at 500 GeV. The Physics Potential}',
Munich, Annecy, Hamburg, ed. P.M.~Zerwas, DESY 92-123 A/B/C,
1992-1993.

\bibitem{ee500new} Proceedings of the Workshop
`{Physics with $e^+e^-$ Colliders}',
Annecy, Gran Sasso, Hamburg, ed. P.M.~Zerwas, DESY 97-100, May 1997.
  
\bibitem{NLC} Proceedings of the Workshop
`{Phy\-sics and Ex\-pe\-ri\-men\-ts with Li\-ne\-ar Col\-li\-ders}',
Sa\-ar\-isel\-k\"a, Fin\-land,
9-14 Sep\-tem\-ber 1991, eds. R.~Orava, P.~Eerola and M.~Nordberg
(World Scientific Publishing, Singapore, 1992).

\bibitem{ppp} Proceedings of the Workshop `Physics and Experiments with 
Linear $e^+e^-$ colliders', eds. F.A.
Harris, S.L. Olsen, S. Pakvasa and X. Tata (World Scientific Publishing,
Singapore, 1993).

\bibitem{LC92} Proceedings of the ECFA workshop on `{$e^+e^-$ Linear
Colliders LC92}', ed. R.~Settles, Garmisch Partenkirchen, July-August
1992, MPI-PhE/93-14, ECFA 93-154, 1993.

\bibitem{JLC} Proceedings of the `I-IV Workshops on Japan Linear Collider',
              KEK, Japan,  1989, 1990, 1992, 1994,
              KEK-Reports 90-2, 91-10, 92-1, 94-1.

\bibitem{sopczak} A. Sopczak, in Ref.~\cite{ee500}, part C.

\bibitem{Kozanecki} W. Kozanecki (convener), in Ref.~\cite{ee500}, part B.

\bibitem{djm} E. Boos, P. Bussey, G. Jikia, D.J. Miller and
J.K. Storrow (conveners), in Ref.~\cite{ee500}, part C.

\bibitem{eph29} J.F. Gunion and H.E. Haber, report UCD-90-25,
1990 (unpublished); in Proceedings of the Summer Study on High
Energy Physics `Research Directions for the Decade', Snowmass, Colorado,
1990, ed. E.L. Berger (World Scientific, Singapore, 1991).

\bibitem{ioeph} S. Moretti, \pr D50 1994 2016.

\bibitem{eph33} D. Bowser-Chao, K. Cheung and S. Thomas, {\it Phys. Lett.} 
{\bf B315} (1993) 399.

\bibitem{HERA} See, e.g.:\\
Proceedings of the Workshop `Future Physics at HERA',
eds. G. Ingelman, A. De Roeck and R. Klanner (DESY, Hamburg, 1995-96).

\bibitem{abu32} I.S. Choi, B.H. Cho, B.R. Kim and R. Rodenberg, \pl B200 
1988 200.

\bibitem{abu33} B. Grzadkowski and W.-S. Hou, \pl B210 1988 233.

\bibitem{abu34} T. Han and C. Liu, \zp C28 1985 295.

\bibitem{gamma_p} G.~Abu Leil and S.~Moretti, {\it
Phys.~Rev.} {\bf D53} (1996) 178.

\bibitem{ankara} {Proceedings of the International Workshop on
`Linac-Ring Type ep and Gamma-p Colliders',} Ankara, Turkey, 9-11 April
1997, to be published in the {\it Turkish Journal of Physics}.

\bibitem{LHC} {Proceedings of the ECFA Large Hadron Collider
Workshop,} Aachen, Germany, edited by G.~Jarlskog and D.~Rein, 
Geneva, Switzerland, 1990, CERN Report
No.~90-10, ECFA Report No.~90-133.

\bibitem{LHC_phys} R.~R\"uckl, in Ref.~\cite{LHC}, Vol.~I.

\bibitem{verdier} A. Verdier, in Ref.~\cite{LHC}, Vol.~III.

\bibitem{LHC_exp} J.~Feltesse, in Ref.~\cite{LHC}, Vol.~I.

\bibitem{Cruz-Sampayo} J.L. Diaz-Cruz and O.A. Sampayo, preprint
UAB-FT-286/92, May 1992 (unpublished).

\bibitem{LHC_top} A.~Ali, F.~Barreiro, J.F.~de Troc\'oniz, G.A.~Schuler
and J.J.~van der Bij, in Ref.~\cite{LHC}, Vol.~II.

\bibitem{epSM} S. Moretti and K. Odagiri, preprint Cavendish-HEP-97/04,
August 1997.



\bibitem{corrMH0iMSSM} Y.~Okada, M.~Yamaguchi and
T.~Yanagida, {\it Prog. Teor. Phys. Lett.} {\bf 85} (1991) 1;\\
J.~Ellis, G.~Ridolfi and F.~Zwirner, {\it Phys. Lett.} {\bf B257} (1991) 83;
{\it Phys. Lett.} {\bf B262} (1991) 477;\\
H.E.~Haber and R.~Hempfling, {\it Phys. Rev. Lett.} {\bf 66} (1991) 1815;\\
R.~Barbieri and M.~Frigeni, {\it Phys. Lett.} {\bf B258} (1991) 395.

\bibitem{corrMHMSSM} A.~Brignole, J.~Ellis, G.~Ridolfi and F.~Zwirner,
                     {\it Phys. Lett.} {\bf B271} (1991) 123;\\
                     A.~Brignole, {\it Phys. Lett.} {\bf B277} (1992) 313;\\
                     H.E.~Haber and M.A.~Diaz, {\it Phys. Rev.}
                     {\bf D45} (1992) 4246.

\bibitem{0pmLEPLHCSSC} V.~Barger, K.~Cheung, R.J.N.~Phillips and A.L.~Stange,
                                  {\it Phys. Rev.} {\bf D46} (1992) 4914.

\bibitem{aroma} G. Ingelman, J. Rathsman and G. A. Schuler,
Comput. Phys. Commun. {\bf 101} (1997) 135 (and references therein).

\bibitem{desy} H1 Collaboration, {\it Z. Phys.} {\bf C74}  (1997) 191;\\
ZEUS Collaboration, {\it Z. Phys.} {\bf C74} (1997) 207.

\bibitem{VEGAS} G.P.~Lepage, {\it Jour.~Comp.~Phys.}~{\bf 27}  (1978) 192.

\bibitem{tim}
T.~Stelzer and W.F.~Long, {\it Comp.~Phys.~Comm.} {\bf 81}  (1994) 357.

\end{thebibliography}
\end{document}